 \documentclass[final,3p,sort&compress,times]{elsarticle}

\usepackage[version=3]{mhchem}

\usepackage{amssymb}
\usepackage{hyperref}
\usepackage{amsmath}
\usepackage{multirow}

\journal{Chemical Engineering Journal}

\begin{document}

\begin{frontmatter}

\title{Poisoning mechanism of ammonia on proton transport and ionomer structure in cathode catalyst layer of PEM fuel cells}

\author[TJU]{Yichao Huang}
\author[TJU]{Zhen Zeng}
\author[TJU,NIEPES]{Tianyou Wang}
\author[TJU,NIEPES]{Zhizhao Che\corref{cor1}}
\cortext[cor1]{Corresponding author. Email: chezhizhao@tju.edu.cn }

\address[TJU]{State Key Laboratory of Engines, Tianjin University, Tianjin, 300350, China.}
\address[NIEPES]{National Industry-Education Platform of Energy Storage, Tianjin University, Tianjin, 300350, China}

\begin{abstract}
Ammonia has strong poisoning effects on cathode catalyst layers of proton exchange membrane (PEM) fuel cells, but the poisoning mechanism is still unclear. In this study, all-atom molecular dynamics simulations are employed to investigate the poisoning mechanisms of ammonia. The results show that ammonium can replace the hydronium ions at the charged sites of sulfonic acid group of the ionomer side chain, and the adsorption of ammonium to sulfonic acid group can be attributed to van der Waals force and electrostatic interaction. Furthermore, other ammonia derivatives, amino and imino ions, can capture hydronium ions to form ion clusters. These ion clusters have strong capability to absorb hydronium ions, and their structures change with ammonia content and temperature. The main mechanism of formation of these clusters is due to the formation of relatively stable hydrogen bonds between ions within the clusters. These mechanisms significantly reduce the efficiency of proton transport, thereby decreasing the catalyst layer's performance in electrochemical reactions. We also discover that the increase in temperature leads to the dissociation of large ion clusters, the blockage in the ionomer layer can be alleviated, and the proton transport efficiency can be restored. The understanding of the poisoning mechanisms obtained in this study is helpful for subsequent research aimed at resolving ammonia poisoning and enhancing the anti-poisoning performance of catalyst layers.
\end{abstract}

\begin{keyword}
\texttt {
Proton exchange membrane fuel cell \sep
Cathode catalyst layer \sep
Ammonia \sep
Proton transport \sep
Molecular dynamics simulation
}
\end{keyword}

\end{frontmatter}

\section{Introduction}\label{sec:1}
Proton exchange membrane fuel cells (PEMFCs) have attracted wide attention in the sustainable development of energy due to their high efficiency and zero emission \cite{daud17, ogungbemi20, peighambardoust10, stambouli11, sun21}. However, the durability of PEMFCs are still key problems that hinder the application and commercialization of PEMFCs \cite{wang22, wei21, yu23}. As the place of electrochemical reaction in PEMFCs, the catalyst layer (CL) can be poisoned by impurity gases such as ammonia, then affecting the stability and performance of fuel cells \cite{hu21, mohtadi04, nagahara08, yang24}. As an efficient hydrogen storage medium, ammonia can enter fuel cells as an impurity during the hydrogen production. Part of the ammonia in the anode can enter the cathode catalyst layer (CCL) in different ways and produce poisoning effects in the CCL \cite{shabani19, zhao22}. Therefore, a thorough understanding of ammonia poisoning mechanism in the catalyst layer is critical for mitigating the harmful influences of ammonia and improving the performance of fuel cells.

Many studies have investigated ammonia as a hydrogen contaminant, focusing on its poisoning effects on the anode of fuel cells. Uribe et al.\ \cite{uribe02} first studied the poisoning of ammonia to fuel cells, introduced ammonia as a fuel contaminant mixed with hydrogen into the fuel cell, and found that trace ammonia would reduce the performance of fuel cells. After ammonia was introduced for a long time, pure hydrogen could not completely restore the performance loss of fuel cells. They used cyclic voltammetry test to measure the catalyst layer after poisoning recovery, and found that the Pt catalyst was not poisoned \cite{uribe02}. Similarly, by introducing ammonia for a period of time and then introducing pure hydrogen, Soto et al.\ \cite{soto03} found that the Pt catalyst was not poisoned by ammonia using the method of cyclic voltammetry test. Therefore, they believe that ammonia has a poisoning effect on the proton exchange membrane by forming ammonium ions to occupy the charged sites in the proton exchange membrane, resulting in a decrease in the proton conductivity. Moreover, ammonia may occupy the active sites on the anode catalyst and hinder the electrochemical reaction. However, the detection method of cyclic voltammetry test uses nitrogen to purge the electrode, which may promote the dissociation of pollutants on the catalyst, so that the peak position changes of current density curve caused by ammonia and its derivatives on the catalyst cannot be observed \cite{jing24}. Haised et al.\ \cite{halseid06} further confirmed that ammonia would have a poisoning effect on the anode catalyst and reduce the progress of hydrogen oxidation reaction (HOR). Imamura et al.\ \cite{imamura09} introduced ammonia at a concentration of 50 ppm into the anode side while simultaneously monitoring the potentials of both the cathode and anode, as well as the emission of ammonia derivatives. They found that ammonia on the anode side diffuses toward the cathode and transfers in the form of ammonium ions, nitrogen oxides, and other compounds. Thus, ammonia can enter the cathode side not only as an air contaminant but also by permeating through the membrane from the anode side. Therefore, the poisoning effects of ammonia on the cathode side also require further investigation.

Many experimental studies have also been conducted on the poisoning effects of ammonia on the cathode side. Damjanovic et al.\ \cite{damjanovic67} studied the key steps of oxygen reduction on platinum catalysts and found that protons significantly influence the rate of oxygen reduction, while ammonia markedly impacts proton activity in the cathode layer, thereby reducing the oxygen reduction reaction (ORR) rate. This phenomenon was also observed by Antoine et al.\ \cite{antoine01}. After studying the effect of ammonia on the HOR at the anode, Haise et al.\ \cite{halseid06} think that ammonia could also hinder the oxygen reduction reaction (ORR) in the cathode catalyst layer. Their subsequent studies have further confirmed this view \cite{halseid04, halseid07, halseid08}. Zhang et al.\ \cite{zhang09} found that ammonia can significantly reduce the electrochemical active surface area (ECSA) of the cathode catalyst layer. Misz et al.\ \cite{misz16} found that ammonia may be oxidized to NO, which in turn occupies the active site on the catalyst. Guo et al.\ \cite{guo15} studied the decomposition of ammonia on platinum catalysts and found that amino and imino ions may also be generated around platinum catalysts. However, there are relatively few studies on the behavior of amino and imino ions in CCLs \cite{wonglakhon22}. Moreover, previous studies on the toxic effect of ammonia on CCLs mostly focused on the concentration of ammonia, the access time of ammonia, and the access time of pure hydrogen \cite{imamura09, jiao21, yoon14, yuan12}, while the effect of temperature on ammonia poisoning was relatively lacking.

Previous studies have shown that ammonia has a significant poisoning effect on both the anode and the cathode catalyst layer through experiments or simulations, and some ammonia poisoning mechanisms have been deduced from experimental phenomena. It is generally accepted that ammonia exerts an irreversible poisoning effect on the ionomer membrane and can also influence electrochemical reactions at both the anode and cathode catalyst layers. However, the mechanisms of the ammonia poisoning effect and the transport behavior of ammonia and its derivatives in the membrane are still unclear. Therefore, in this study, we focus on the effect of ammonia in the CCL on proton transport in the ionomer layer. Molecular dynamics simulation is performed by not only adding ammonia to the CCL, but also adding ammonium, amino and imino ions as ammonia derivatives. Different ammonia contents and temperatures are set to further study the poisoning mechanism of ammonia in the CCL at the molecular scale. The behavior of amino and imino ions capturing hydronium ions to form ion clusters in the CCL is explained, and the structure and distribution characteristics of ion clusters are analyzed by using the radical distribution function (RDF), coordination number (CN), cluster analysis, and hydrogen bond analysis. The pore size distribution (PSD) and hydrogen bond analysis are also used to explore the effect of ammonia on the structure of the ionomer layer and the adsorption mechanism. RDF, CN and hydrogen bond analysis are also used to explore the effect of ammonia on the structure and distribution of water clusters.

\section{Methods}\label{sec:2}
\subsection{Molecular models}\label{sec:2.1}
In the CCL model, the carbon substrate consists of five layers of carbon atoms, each layer contains 1344 carbon atoms, and the interval between each layer is 3.354 \AA. The dimensions of the simulation box in the $x$-$y$ directions are 58  $\times$ 59 \AA$^2$ with periodic boundary conditions. The $z$-dimension was set to very large (i.e., 350 \AA) to avoid the influence of the periodic boundary condition. A truncated-octahedral platinum particle was placed on the carbon substrate. Previous studies have pointed out that this truncated-octahedral shape of platinum particle is the most consistent with platinum particles in actual PEMFCs \cite{cheng10}. The platinum particle contains 586 platinum atoms, and the size is about 2.35 \AA. The Pt/C substrate was rigid and fixed in the whole simulation process, because the shape change of the substrate is not the focus of the study. Nafion was adopted as the PFSA ionomer in this model, each PFSA ionomer contains ten repeating units and each unit contains a sulfonic acid group ($\ce{SO3-}$) \cite{damasceno13}. The equilibrium weight (EW) of the ionomer is 1144 g/mol. The water content ($\lambda$) is defined as the number of water molecules and hydronium ions per sulfonic acid group,
\begin{equation}\label{eq:01}
  \lambda ={\left( {{N}_{{{\text{H}}_{\text{2}}}\text{O}}}+{{N}_{{{\text{H}}_{\text{3}}}{{\text{O}}^{\text{+}}}}} \right)}/{{{N}_{\text{SO}_{3}^{-}}}}
\end{equation}
where ${{N}_{{{\text{H}}_{\text{2}}}\text{O}}}$, ${{N}_{{{\text{H}}_{\text{3}}}{{\text{O}}^{\text{+}}}}}$ and ${{N}_{\text{SO}_{3}^{-}}}$ are the number of water molecules, hydronium ions and sulfonic acid groups. The ionomer mass content ($\varphi $) is defined as
\begin{equation}\label{eq:02}
  \varphi ={{{m}_{\text{ionomer}}}}/{\left( {{m}_{\text{ionomer}}}+{{m}_{\text{C-plate}}}+{{m}_{\text{Pt-particle}}} \right)}
\end{equation}
where ${{m}_{\text{ionomer}}}$, ${{m}_{\text{C-plate}}}$, and ${{m}_{\text{Pt-particle}}}$ are the mass of ionomer, carbon substrate and platinum particle, respectively. The water content ($\lambda$) was set at 16, which has been considered to be the optimal operating condition of PEMFCs \cite{chen20}. The ionomer mass content was set to 31.9\%, at which value the ionomer can be fully hydrated without agglomeration \cite{huang24}. Therefore, 1200 water molecules and 8 PFSA ionomers were placed on the Pt/C substrate, and 80 hydronium ions were added to maintain the electrical neutrality. The number of oxygen molecules was fixed at 80. To study the transport and distribution of ammonia and its derivatives in the CCL model, ammonia ($\text{NH}_{3}^{{}}$) molecules, ammonium ($\text{NH}_{4}^{+}$), amino ($\text{NH}_{2}^{-}$) and imino ($\text{NH}_{{}}^{2-}$) ions were randomly placed in the system. To simulate the structure and transport characteristics of the CCL at different ammonia concentrations, different ammonia contents ($\psi $) were set, which is defined as
\begin{equation}\label{eq:03}
  \psi ={\left( {{N}_{\text{ammonia}}}+{{N}_{\text{ammonium}}}+{{N}_{\text{amino}}}+{{N}_{\text{imino}}} \right)}/{{{N}_{\text{SO}_{3}^{-}}}}
\end{equation}
where ${{N}_{\text{ammonia}}}$ is the number of ammonia molecules ($\ce{NH3}$), ${{N}_{\text{ammonium}}}$, ${{N}_{\text{amino}}}$ and ${{N}_{\text{imino}}}$ are the number of ammonium ($\ce{NH4+}$), amino ($\ce{NH2-}$), and imino ($\ce{NH}^{2-}$) ions, respectively, and ${{N}_{\text{SO}_{3}^{-}}}$ is the number of sulfonic acid groups. The ammonia content was varied from $\psi $ = 0 to 4 in this study. To maintain the electrical neutrality of system, the ratio between ammonia and its derivatives is set to
  ${{N}_{\text{ammonia}}}:{{N}_{\text{ammonium}}}:{{N}_{\text{amino}}}:{{N}_{\text{imino}}}=3:4:2:1$.
To analyze the effect of temperature on the ammonia toxicity, different system temperatures were considered, i.e., 298, 333, 343, 353, and 363 K.

\subsection{Simulation details}\label{sec:2.2}
In this study, F3C and classical hydronium models were used to describe water molecules and hydronium ions \cite{michael97, seung04}. The classical DREIDING force field was used for oxygen, ammonia, ammonium, amino and imino ions \cite{stephen90}. The modified DREIDING force field was used for PFSA ionomers \cite{mabuchi14}. The parameters developed by He et al.\ \cite{he13} was used for the force fields of the platinum and carbon atoms in the Pt/C substrate with other atoms of the system. The interaction between different types of atoms follows the Lorentz-Bertelot rule. The atomic charges of the ammonia, ammonium, amino, and imino were calculated using Mulliken population analysis with the double numerical basis set with polarization (DNP) and GGA-PBE functional. The cut-off distance of Lennard-Jones (LJ) interaction was set as 1.5 nm. The long-range electrostatic interactions were calculated by the particle-particle particle-mesh (PPPM) method with an accuracy of 0.0001. The Nos\'{e}-Hoover thermostat with relaxation time of 0.1 ps was used to control the system temperature. All the MD simulations were performed using the LAMMPS package \cite{plimpton95} with a timestep of 1 fs.

\begin{figure}
  \centering
  \includegraphics[scale=0.6]{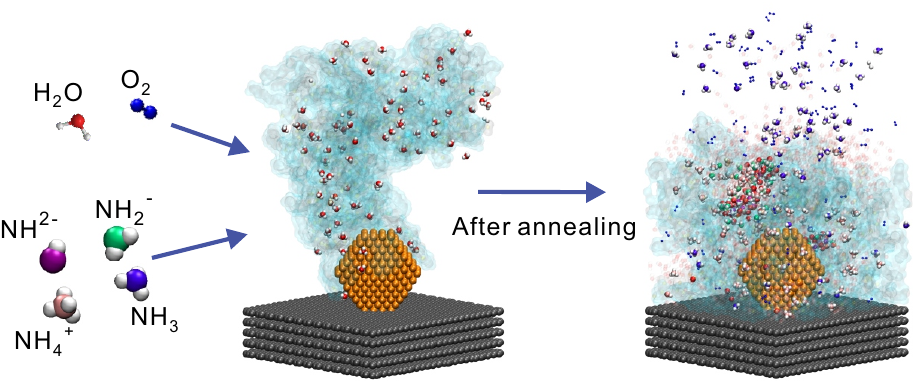}
  \caption{Configurations of CCL model before and after the annealing process.}\label{fig:01}
\end{figure}

After the ionomer and hydronium ions were placed on the Pt/C substrate, an NVT simulation of 2 ns at 298 K was performed to make the initial relaxation of the ionomer, so that the ionomer can spread on the Pt/C substrate. Then water and oxygen molecules, as well as ammonia and its derivatives, were randomly placed in the ionomer to start an annealing process, which is described in our previous study \cite{ huang24}. The structural changes of the system during annealing process are shown in Figure \ref{fig:01}. During the annealing process, a fixed wall (9 nm from the Pt/C substrate) was used to prevent the water molecules from escaping by excessive evaporation. Then another NVT simulation for 26 ns at 353K was performed. The last 10 ns of the NVT simulation was used for data sampling. At this time, the height of the fixed wall was increased to 13 nm from the Pt/C substrate to avoid any confinement effect. The parameters developed by Fan et al.\ \cite{fan19, fan20} was adopted for the potential of the fixed wall.

\section{Results and discussion}\label{sec:3}
\subsection{Equilibrated structures of the CCL with ammonia poisoning}\label{sec:3.1}
\begin{figure}
  \centering
  \includegraphics[scale=0.5]{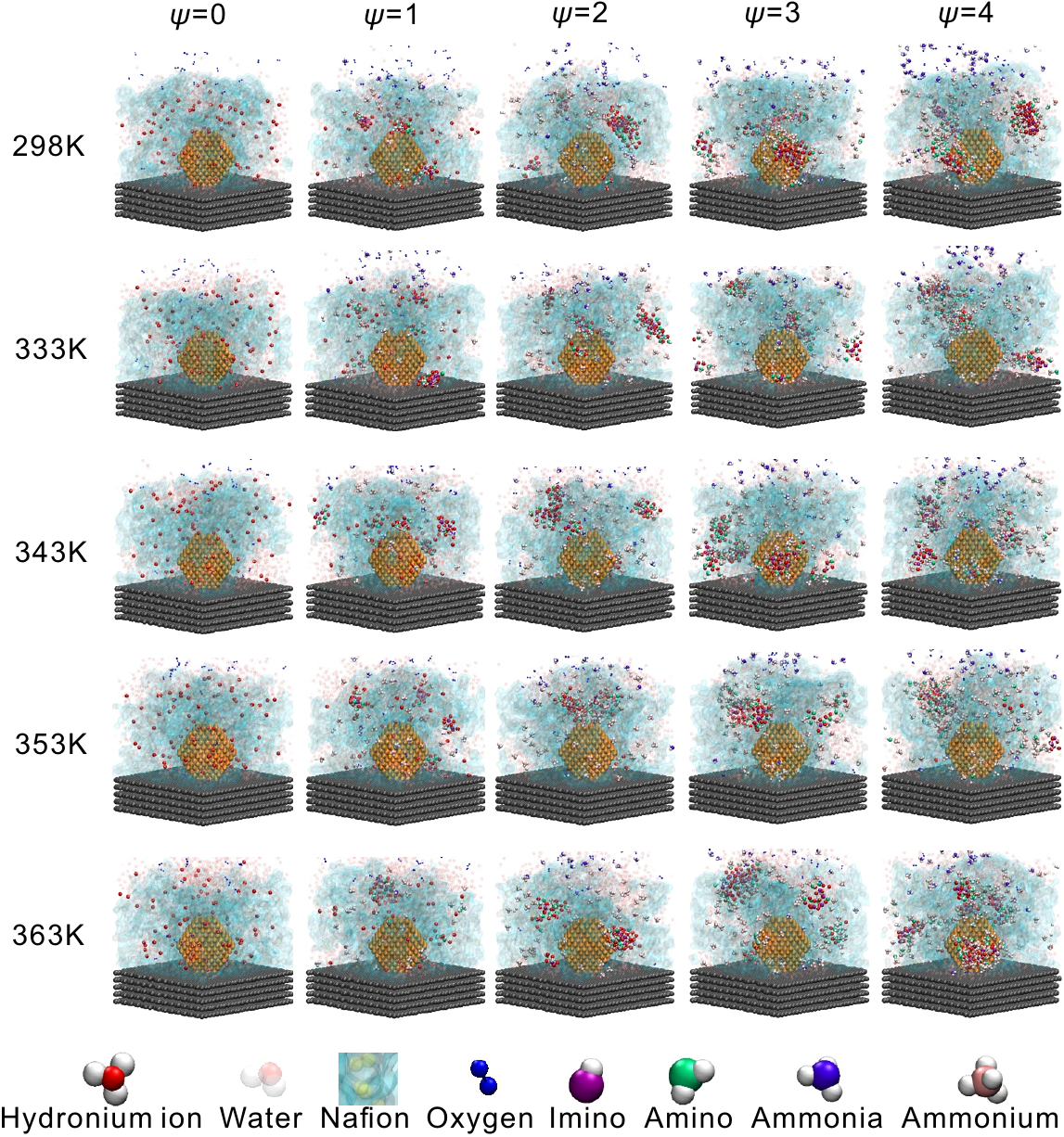}
  \caption{Snapshots of equilibrated systems with different ammonia contents ($\psi $) and different system temperatures.}\label{fig:02}
\end{figure}

In this study, the CCL model includes a carbon-supported platinum (Pt/C) substrate, perfluoro-sulfonic acid (PFSA) ionomers, water, hydronium ions, and oxygen molecules, as shown in Figure \ref{fig:01}. The snapshots of CCLs at the end of the simulations at different ammonia contents and different temperatures are shown in Figure \ref{fig:02}. The water molecules and ionomers are shown in transparent to highlight the structure of ammonia and its derivatives in the ionomer. When there is no ammonia (i.e., $\psi $ = 0), the hydronium ions are evenly distributed in the ionomer, indicating that the ionomer layer is conductive to the transport of protons to the catalyst surface for electrochemical reactions. After adding a small amount of ammonia to the system ($\psi $ = 1), we can see that some ammonia is distributed in and above the ionomer, while some ammonium ions occupy the original position of hydronium ions, i.e., charged sites of the sulfonic acid groups on the side chain of the ionomer. The amino and imino ions are combined with the hydronium ions to form some small ion clusters, which can transfer through pores inside the ionomer. After adding more ammonia into the system ($\psi $ = 2, 3), we can see that all free hydronium ions are captured by amino and imino ions to form larger ion clusters. These large ion clusters are blocked in the pores of the ionomer, and their positions are relatively fixed, which will greatly increase their transport resistance in the CCL. When the ammonia content is high ($\psi $ = 4), several large ion clusters are formed at different positions in the ionomer layer. The charged sites of sulfonic acid group are mostly occupied by the ammonium ions instead of hydronium ions. Some free amino and imino ions are distributed in the ionomer layer, indicating that they can capture more hydronium ions. In this condition, it is difficult for hydronium ions to reach the surface of the platinum catalyst, resulting in a great increase in the proton transport resistant.

\subsection{Structural and distribution analysis of ammonia ion clusters}\label{sec:3.2}
From the snapshots of CCLs shown in Figure \ref{fig:02}, we can see that amino and imino ions can capture many hydronium ions to form ion clusters. Since the presence of ion clusters has serious poisoning effects on the ionomer, the structure of the ion clusters is analyzed here. We characterize the structure of the ion clusters using the radical distribution function (RDF) ${{g}_{A-B}}(r)$
\begin{equation}\label{eq:05}
  {{g}_{A-B}}(r)={\left( \frac{{{n}_{B}}}{4\pi {{r}^{2}}dr} \right)}/{\left( \frac{{{N}_{B}}}{V} \right)}
\end{equation}
where ${{n}_{B}}$ is the number of atoms B located at a distance $r$ in a shell of thickness $dr$ from atom A, ${{N}_{B}}$ is the number of atoms B in the system and $V$ is the volume of the system. The ion clusters contain amino, imino, and hydronium ions. Hence, the RDF analysis was performed on each of them in the form of ${{g}_{\text{NH-OI}}}$, ${{g}_{\text{N}{{\text{H}}_{\text{2}}}\text{-OI}}}$ and ${{g}_{\text{NH-N}{{\text{H}}_{\text{2}}}}}$, where OI represents the oxygen atom in the hydronium ions, NH and \ce{NH2} represents the nitrogen atom in imino and amino ions. The coordination numbers (CN) of these pairs were also calculated by integrating the RDF within the first solvation shell.

\begin{figure}
  \centering
  \includegraphics[scale=0.75]{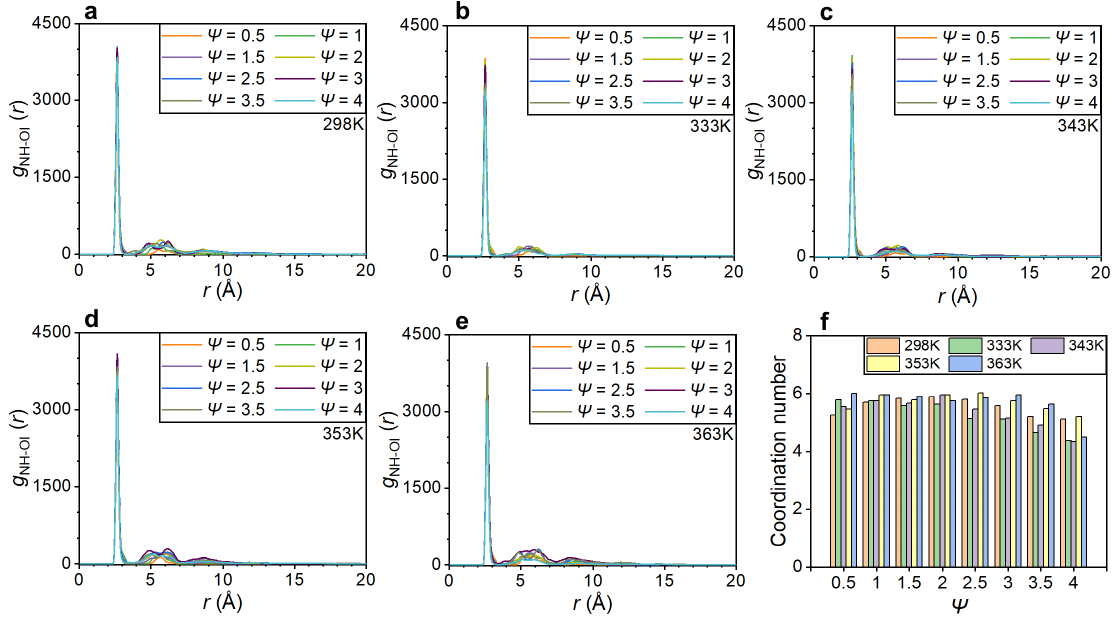}
  \caption{(a-e) RDFs of NH-OI at various ammonia contents and temperatures. (f) CN of NH-OI. }\label{fig:03}
\end{figure}

\begin{figure}
  \centering
  \includegraphics[scale=0.75]{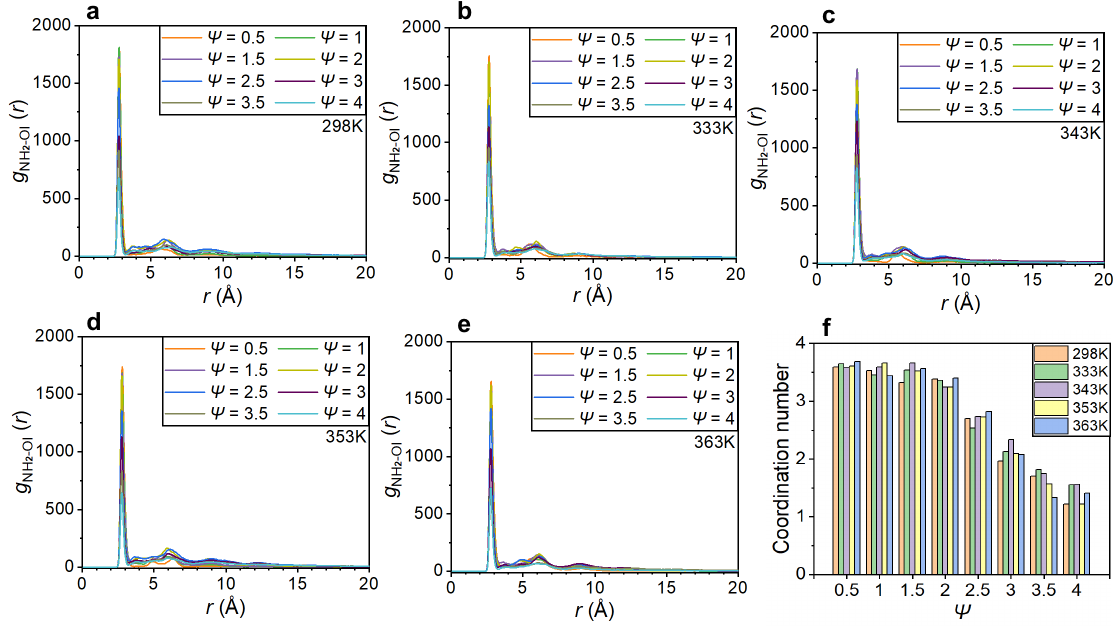}
  \caption{(a-e) RDFs of \ce{NH2}-OI at various ammonia contents and temperatures. (f) CN of \ce{NH2}-OI.}\label{fig:04}
\end{figure}

\begin{figure}
  \centering
  \includegraphics[scale=0.75]{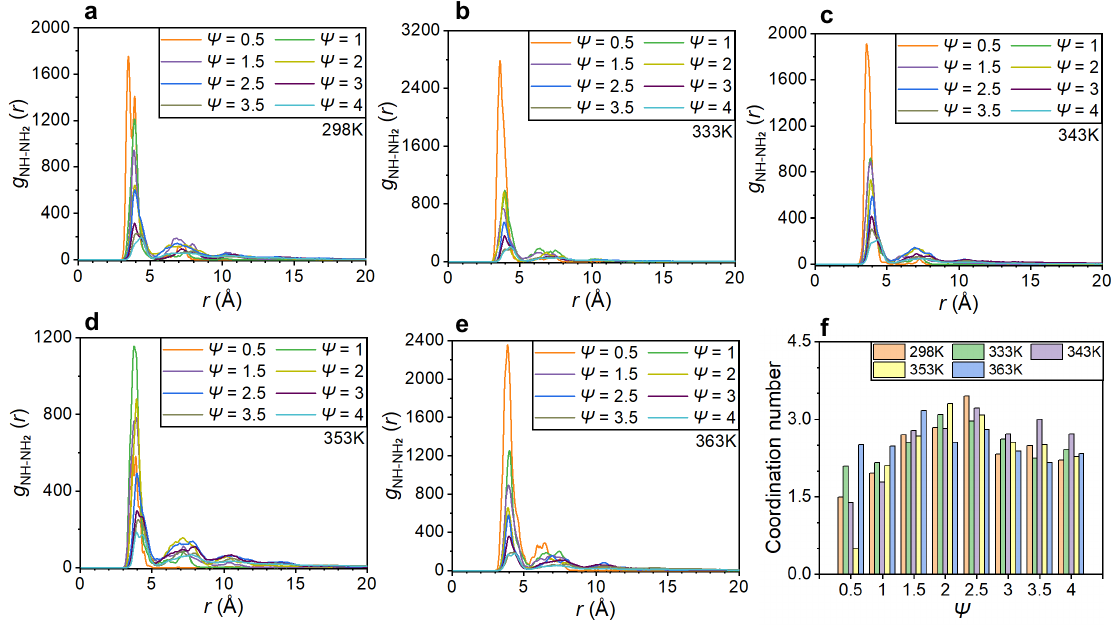}
  \caption{(a-e) RDFs of NH-\ce{NH2} at various ammonia contents and temperatures. (f) CN of NH-\ce{NH2}.}\label{fig:05}
\end{figure}

The RDFs of ${{g}_{\text{NH-OI}}}$, ${{g}_{\text{N}{{\text{H}}_{\text{2}}}\text{-OI}}}$ and ${{g}_{\text{NH-N}{{\text{H}}_{\text{2}}}}}$ and the CNs are shown in Figures \ref{fig:03}--\ref{fig:05}, respectively. In Figure \ref{fig:03}, we can find that the first peak of the RDF of the NH-OI pair appears at about 2.65 \AA, and the first valley appears at about 3.45 \AA. The position of the first valley was used as the radius of the first solvation shell to integrate the RDF in the calculation of the CN, and the result is shown in Figure \ref{fig:03}(f). We can see that the correlation between imino and hydronium ions is subtly affected by the system temperature, while the CN decreases from about 5.99 to 4.34 as the ammonia content increases from 0.5 to 4. This is because, with the increase of ammonia content, the average number of imino in the ion clusters also increases, resulting a decrease in the CN.

The RDFs and CNs of the \ce{NH2}-OI pair are shown in Figure \ref{fig:04}. The first peak of the RDF appears at about 2.75 \AA, and the first valley also appears at about 3.45 \AA, which is the radius of the first solvation shell. In Figure \ref{fig:04}(f), we can find that the effect of temperature on the correlation between amino and hydronium ions is still very slight, while the effect of the ammonia content is significant. As the ammonia content increases from 0.5 to 4, the CN decreases sharply from 3.66 to 1.21. This is because with the increase in the ammonia content, all hydronium ions have been captured by the ion clusters, while the number of amino ions still increases, hence, the CN decreases significantly. In addition, because the negative charges carried by amino ions is less than the imino ions, the CNs of \ce{NH2}-OI is less than that of NH-OI pair. Moreover, the stability of imino ions to the captured hydronium ions is higher than that of amino ions, hence the change in the CN caused by the change in the ammonia content is much smaller. However, if only from the perspective of charge absorption, one imino ion can only capture two hydronium ions and one amino ion can only capture one hydronium ion, which is inconsistent with the CN results. Therefore, we further analyze the RDF and CN between amino and imino ions, as shown in Figure \ref{fig:05}. In Figure \ref{fig:05}, we can see that the position of the first peak appears at about 3.75 to 3.95 \AA, and the first valley appears at about 5.15 \AA. From the CN, we can see that at low ammonia contents, the CN fluctuation affected by temperature changes is larger than that at high ammonia contents. Because at low ammonia contents, the sizes of the ion clusters are relatively small, and they can move through the pores in the ionomer layer, and will also be affected by the charged sites from the ionomer side chains. Therefore, the change in the molecular activity caused by temperature change will significantly affect the stability of ion clusters, resulting in the change in the CN. When the ammonia content increases, the ion clusters will grow larger and be blocked in the pores of the ionomer layer. In this condition, the stability of ion clusters is higher, so the change in the CN caused by temperature change is not obvious. In Figure \ref{fig:05}(f), we can see that the CN of NH and \ce{NH2} increases first and then decreases with the increase in the ammonia content. The CN reaches the highest value when the ammonia content is 2.5, indicating that the ion clusters have the highest stability in this condition, and it will also be easier to form larger ion clusters. These large clusters will occupy the space inside the ionomer layer, even affect the morphology of the ionomer layer. When the hydronium ions go through the pores and reaches these spaces, it will be difficult for them to escape from the attraction of ion clusters, thereby hindering the electrochemical reaction. This indicates that ammonia and its derivatives severely hinder the transport of hydronium ions, thereby affecting the oxygen reduction reaction in the cathode catalyst layer, being consistent with previous experimental studies \cite{halseid04, halseid07, halseid08}. When the ammonia content continues to increase, the ion clusters tend to be stable, and the other amino and imino ions will not be absorbed by these ion clusters. Therefore, it will generate some small ion clusters, resulting in a decrease of the CN.

Through the above results of RDFs and CNs, we can see that some general structural characteristics of ion clusters. The imino ions are stable because they are at the core of the ion clusters; while the amino ions are often affected by the change in temperature and ammonia content because they repeatedly detach and attach at the outer layer of the ion clusters. The amino ions and imino ions have the maximum correlation when the ammonia content is 2.5, indicating that the stability of the ion cluster structure and the strong capability of adsorbing hydronium ions in this condition.

\begin{figure}
  \centering
  \includegraphics[scale=0.75]{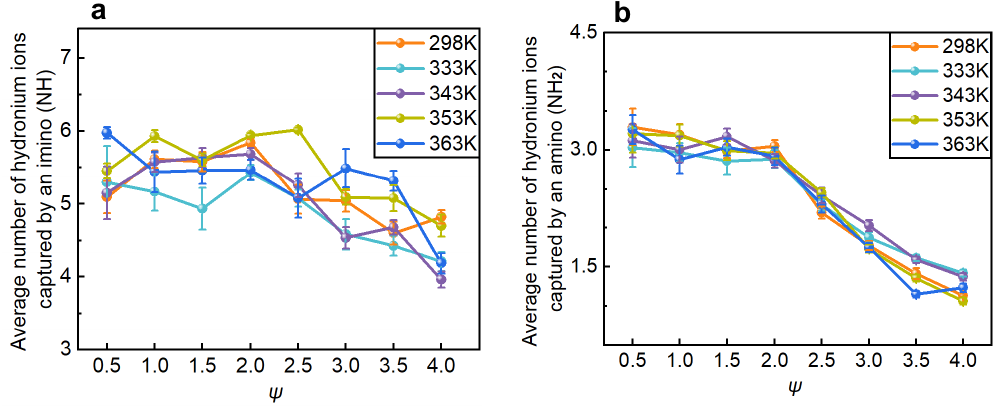}
  \caption{Average number of hydronium ions captured by (a) an imino ion or (b) an amino ion at different ammonia contents ($\psi $) and temperatures.}\label{fig:06}
\end{figure}

To further analyze the interaction between ions in the ion clusters, we calculate the average numbers of hydrogen bonds between imino and hydronium ions, and between amino and hydronium ions. By dividing the number of amino and imino ions, we can calculate the numbers of hydronium ions captured by amino and imino ions through hydrogen bond forces. The oxygen atoms of hydronium ions are set as donors, and the nitrogen atoms in imino and amino ions are set as acceptors. When the distance between the donors and the acceptors is less than 3.5 \AA\, and the angle of donor-hydrogen-acceptor is less than 41.7$^\circ$, we determine that hydrogen bonds are formed. In the calculation, 200 configurations of the last 10-ns NVT simulation are used, and the average results are shown in Figure \ref{fig:06}. All calculations of hydrogen bonds are based on the MDAnalysis package \cite{gowers16, michaud-agrawal11}. As shown in Figure \ref{fig:06}, the number of hydronium ions that can be captured by amino and imino ions with the change of ammonia contents and temperatures is similar to the results of the CN analysis discussed previously. This is mainly because the determination distance of hydrogen bond formation is 3.5 \AA, while the radius of the first solvation shell used to calculate the CN is 3.45 \AA. Since the analysis of hydrogen bond adds the angle determination, we believe that the strong adsorption of amino and imino ions to hydronium ions mainly depends on the formation of hydrogen bonds, and the electrostatic interaction is not the dominant factor in this condition. In Figure \ref{fig:06}, we can also find that with increasing the ammonia content, the ability of amino and imino ions to capture hydronium ions through hydrogen bonds decreases, while imino ions are more susceptible to temperature changes than amino ions, resulting in stronger fluctuations in the average number of hydronium ions captured by an imino. This is because the larger negative charge value of the imino ion will attract the hydronium ions that are further away and repeatedly form hydrogen bonds with the imino ions. The repeated formation of these hydrogen bonds is mainly affected by the molecular activity, which is mainly influenced by the temperature. Moreover, we compare the CN of amino and hydronium ions with the number of hydrogen bonds formed, and the results show that the number of formed hydrogen bonds is slightly larger than the CN. This is because the hydronium ions attracted by the imino ion will also form hydrogen bonds with the amino ion which is in the outer layer of the ion clusters, resulting in an increase in the number of hydrogen bonds. Furthermore, the structural characteristics of ion clusters can be obtained by calculating the RDF and CN of amino and imino ions with water molecules. As shown in Figure S1 in Supplementary Material, the coordination number between amino ions and water molecules is significantly higher than that between imino ions and water molecules, indicating that the amino ions are in the outer layer of the ion cluster, and the imino ions are surrounded by the amino ions and are in the inner layer of the ion cluster.

\begin{figure}
  \centering
  \includegraphics[scale=0.65]{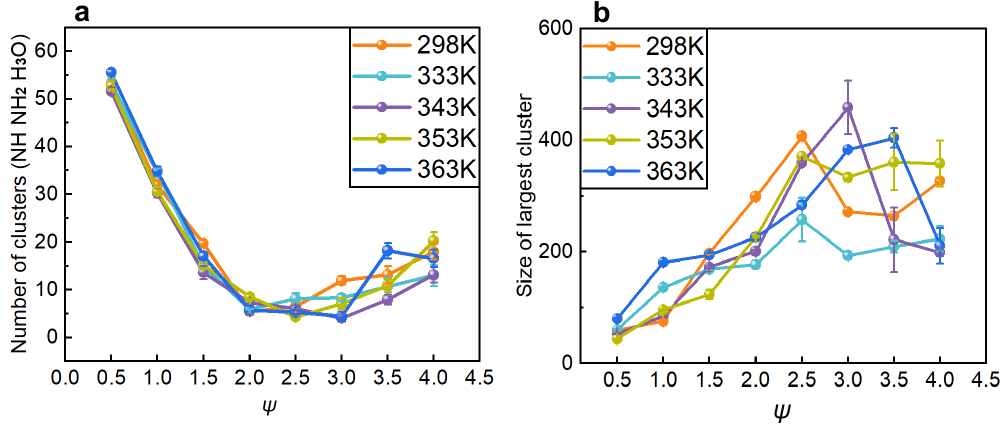}
  \caption{(a) Average number of ion clusters at different ammonia contents ($\psi $) and temperatures. (b) Size of largest ion clusters at different ammonia contents ($\psi $) and temperatures.}\label{fig:07}
\end{figure}

To further analyze the distribution of the ion clusters in the ionomer layer, we performed cluster analysis of the ion clusters. We counted the number of clusters and calculated the sizes of the largest ion clusters in various conditions. As shown in Figure \ref{fig:07}(a), the number of ion clusters is not affected by the change of the system temperature, and is mainly affected by the ammonia content. With the increase in the ammonia content, the number of ion clusters decreases first and then increases. This is because when the ammonia content is low, the hydronium ions are mostly in a free state, and does not form ion clusters. With the increase in the ammonia content, more hydronium ions are captured by amino and imino ions, leading to the decrease in the number of clusters. After all hydronium ions are captured, further adding ammonia will make some amino and imino ions in the free state, which increases the number of ion clusters. In Figure \ref{fig:07}(b), we can see that the size of the largest ion cluster increases with the increase in the ammonia content. When the ammonia content reaches 2.5, the size of the ion cluster reaches the maximum, while the effect of temperature changes is still subtle.

\begin{figure}
  \centering
  \includegraphics[scale=0.65]{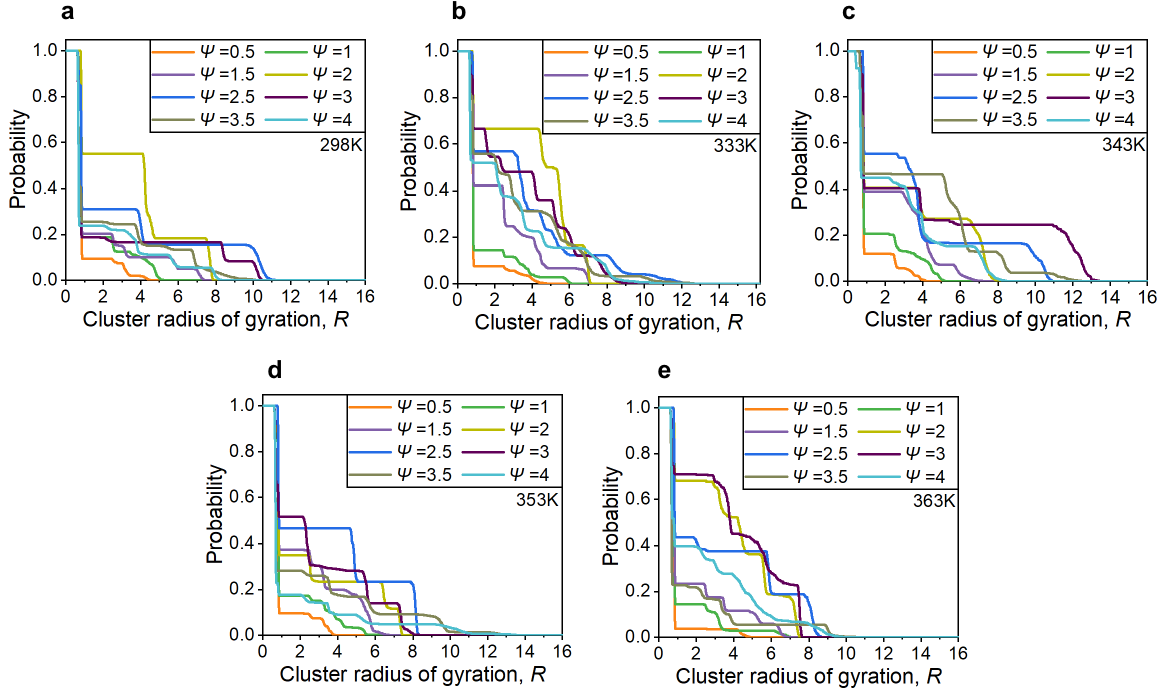}
  \caption{Complementary cumulative distribution function of the radius of gyration $R$, i.e., the probability of finding an ion cluster with the radius of gyration greater than or equal to $R$.}\label{fig:08}
\end{figure}

Since the number and size of ion clusters alone cannot reflect the degree of blockage of ion clusters within the ionomer layer, we calculated the radius of gyration of the ion clusters, which can reflect the difficulty of ion clusters transport in the ionomer layer. Figure \ref{fig:08} shows the complementary cumulative distribution function of the radius of gyration $R$, i.e., the probability of finding an ion cluster with the radius of gyration larger than or equal to $R$. The analysis is based on the average results obtained from the last 10-ns NVT simulations. The cluster analysis was performed in the OVITO software \cite{stukowski10}, and the probe radius was set to 3.5 \AA. To obtain an accurate distribution of the radius of gyrations, we set the bin size of calculating the probability with respect to $R$ to 0.01 \AA. In Figure \ref{fig:08}, we can see that the radius of gyration is mainly distributed between 1--8 \AA, and the probability of finding ion clusters larger than this range is less than 0.3. We can also see that larger radius of gyration mainly occurs when the ammonia content is greater than or equal to 2.5. In this condition, the correlation between ion clusters has reached its maximum value according to the above analysis of RDF and hydrogen bonds. At low temperatures (e.g., 298 and 333 K) and when the ammonia content reaches 2.5, the number of particles contained in ion clusters reaches its maximum value, and the molecular activity is relatively low. Free amino and imine ions are distributed in other pores within the ionomer layer, making it difficult to easily reach the space of these large ion clusters. However, as the temperature increases (e.g., 343 K), the activity of free amino and imine ions increases, and with a higher probability of reaching the space of the original large ion clusters, they will be adsorbed by the outer amino-hydronium ion structure, leading to a further increase in the cluster's radius of gyration, explaining the occurrence of some clusters whose sizes reach 13--14 \AA. As the temperature further increases and the molecular activity further increases, the stability of the outer layer of large ion clusters is affected. Some larger ion clusters dissociate into medium-sized ion clusters, resulting in a decrease in the radius of gyration of ion clusters at 363 K. This reduction may be beneficial for the transport of ion clusters in the ionomer layer, reducing the negative impact of ion cluster blockage.

\subsection{Ionomer structure and ionomer-ammonia interaction}\label{sec:3.3}
At the optimal working condition of catalyst layers, water and hydronium ions will transfer in the micropores of the hydrated ionomer layer. After adding ammonia into the system, some large ion clusters form in the ionomer layer. To analyze how the addition of ammonia affects the structure of the ionomer layer, we performed the pore size distribution (PSD) analysis of the hydrated ionomer layer, which can well reflect the pore state in the ionomer. In the pore size distribution analysis, we removed the water molecules, hydronium ions, ammonia and its derivatives in the hydrated ionomer layer. Moreover, because the platinum particle is wrapped by the ionomer, the inclusion of platinum in the PSD analysis can better reflect the pore distribution of ionomer layer. All pore size distributions were calculated using the zeo++ software package \cite{pinheiro13, willems12}, and the probe radius was set to 1.4 \AA. The PSD data of 200 configurations in the last 10 ns sampling stage are averaged to obtain the final PSD results.

\begin{figure}
  \centering
  \includegraphics[scale=0.58]{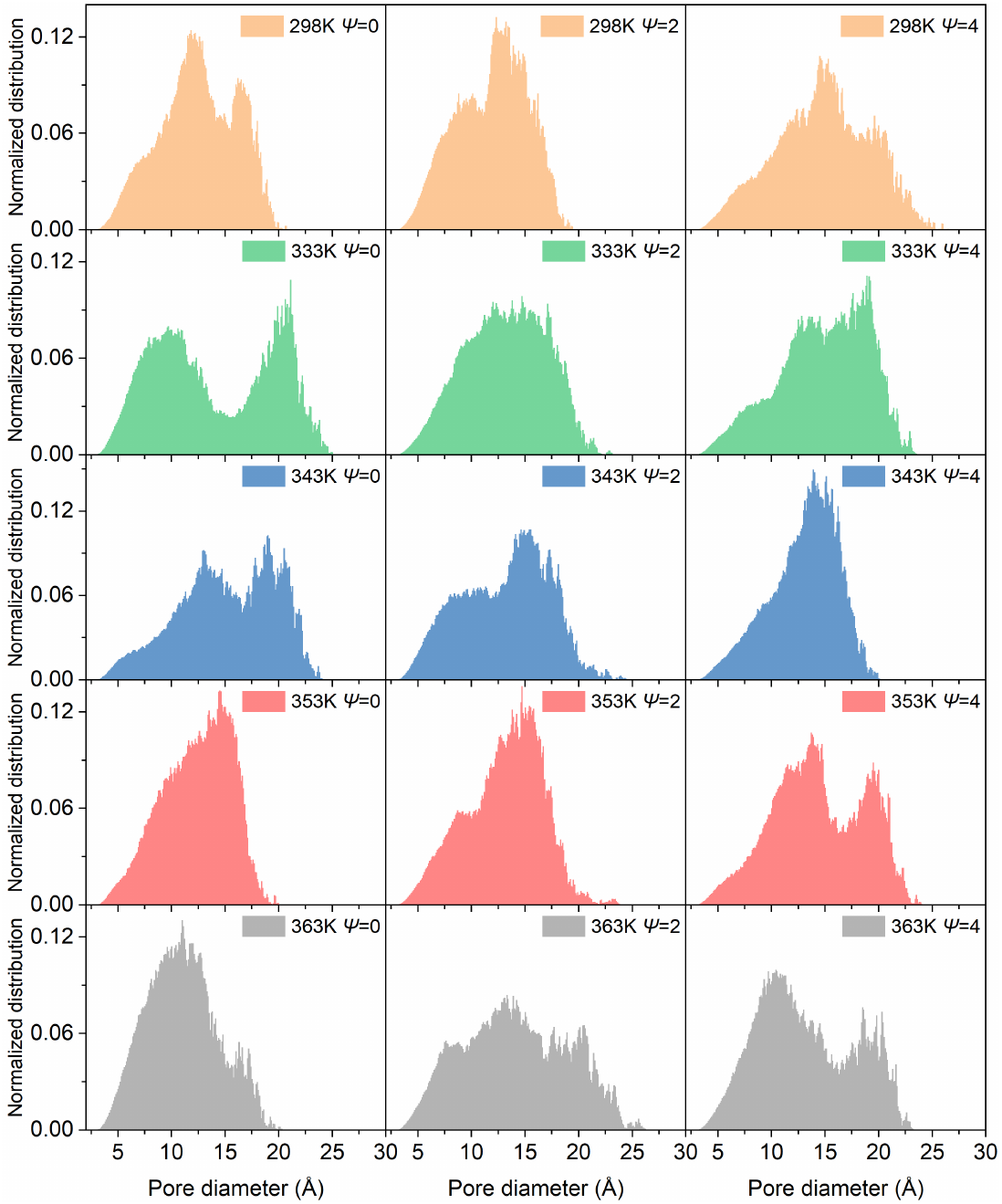}
  \caption{Pore size distribution of PFSA ionomer layer after removing water clusters and ion clusters at various temperature and ammonia contents.}\label{fig:09}
\end{figure}

From Figure \ref{fig:09}, we can find that the addition of ammonia will have a great impact on the pore size distribution of ionomer. At low temperatures (298, 333, 343 K), the pore size distribution of the ionomer changes from bimodal to unimodal. The structure of the ionomer has changed from a relatively loose and porous structure to a structure of one single large hole and multiple small holes, and the uniformity of the ionomer layer has been further reduced. We can also find that with increasing the ammonia content, the size of the pores that are most likely to appear in the ionomer layer also increases accordingly. These results suggest that the ion clusters are mainly blocked in these large pores and cannot diffuse into other pores. Comparing the size of large pores in the ionomer layer with the radius of gyration of ion clusters under the corresponding conditions, we can find that the pore sizes are larger, which indicates that there are also a number of water molecules inside these pores. These water molecules can provide channels for proton transport. However, when protons reach the space with large ion clusters, the protons will be attracted strongly and cannot escape. Under these conditions, the transport resistance in the ionomer layer will be greatly increased. While when the temperature is high (e.g., 353 or 363 K), the pore size distribution curve will change from unimodal to bimodal with the increase in the ammonia content, and the trend is opposite to the results at low temperatures. However, the changes here is in line with expectations. From the previous analysis of ion clusters, the large-sized ion clusters will be decomposed into middle-sized ion clusters under high temperatures. Multiple ion clusters may exist in one pore to form a huge hole inside the ionomer layer, or may be distributed in different pores forming a number of large holes, which leads to the formation of bimodal in the pore size distribution curve. Moreover, due to the existence of multiple ion clusters, the ionomer structure is squeezed in different directions, and the homogeneity of the ionomer is improved. The improvement of the homogeneity will be beneficial to the transport of hydronium ions in the region without ion clusters, which can effectively reduce the transport resistance in the ionomer layer.

To reflect the blockage of ammonia ion clusters in the ionomer layer intuitively, we calculated the mean square displacement (MSD) of hydronium ions at different temperatures and ammonia contents, as shown in Figure S2 in Supplementary Material. We can see that at different temperatures, with the increase in the ammonia content, the MSD of hydronium ions decrease significantly. When the ammonia content reaches 2, with the formation of stable ion clusters, the MSD value of hydronium ions is very small, indicating that the movement of ion clusters is limited. This further reflects that the ion clusters are blocked in the ionomer layer. We can also see that when the temperature reaches 363 K, the MSD of hydronium ions has a larger value than at lower temperatures, suggesting that increasing the temperature can effectively alleviate the blockage state in the ionomer layer and accelerate the transport of components.

\begin{figure}
  \centering
  \includegraphics[scale=0.55]{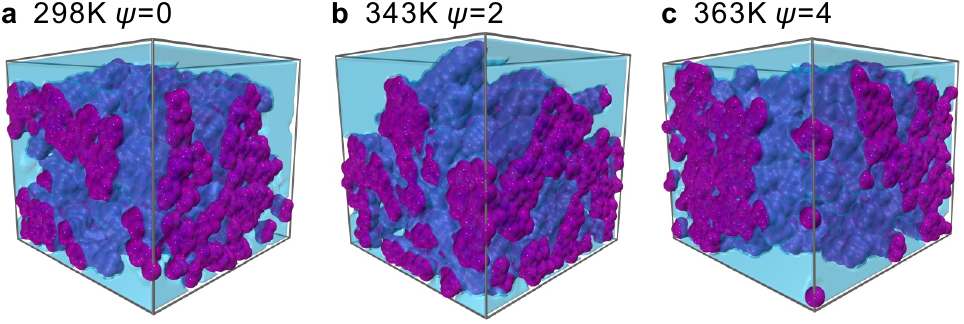}
  \caption{Snapshots of ionomer layer and free space at different working conditions. Purple ball means the ionomer, and the cyan transparent part means free spaces inside the ionomer layer.}\label{fig:10}
\end{figure}

To further analyze the influence of the change of ammonia content on the internal pore structure of the ionomer, we show the structures of the ionomer pores in three working conditions, namely low temperature without ammonia, medium temperature with medium ammonia content, and high temperature with high ammonia content, as shown in Figure \ref{fig:10}. The snapshots can intuitively reflect the information presented in the pore size distribution curve in Figure \ref{fig:10}, demonstrating the influence of the existence of ion clusters on the structure of the ionomer layer. From the snapshots of ionomer and free region at low temperature without ammonia (298 K, $\psi $ = 0 in Figure \ref{fig:10}(a)), it can be found that the hydrated ionomer is homogeneous, and there are sufficient free spaces on the side, top, and bottom of the ionomer layer. The large free space is conducive to the diffusion of water and hydronium ions. At medium temperature with medium ammonia content as shown in Figure \ref{fig:10}(b), we can see that the free space at the bottom and side of the ionomer layer is reduced, the ionomer molecules at the top are stretched, and the homogeneity of the ionomer layer is significantly reduced. This is because the internal ion clusters squeeze the ionomer space and make it expand outward. The influence of ammonia ion clusters on the membrane structure also explains the previously observed experimental phenomenon of membrane mass transfer performance being affected by ammonia poisoning \cite{yoon14, misz16}. Figure \ref{fig:10}(c) shows the state of the ionomer at high temperature and high ammonia content. Because the multiple ion clusters squeeze the ionomer in multiple directions, the ionomer expands in all directions, the free space at the top and bottom of the ionomer also increases accordingly, and the homogeneity increases again, providing space for water and proton transport.

\begin{figure}
  \centering
  \includegraphics[scale=0.8]{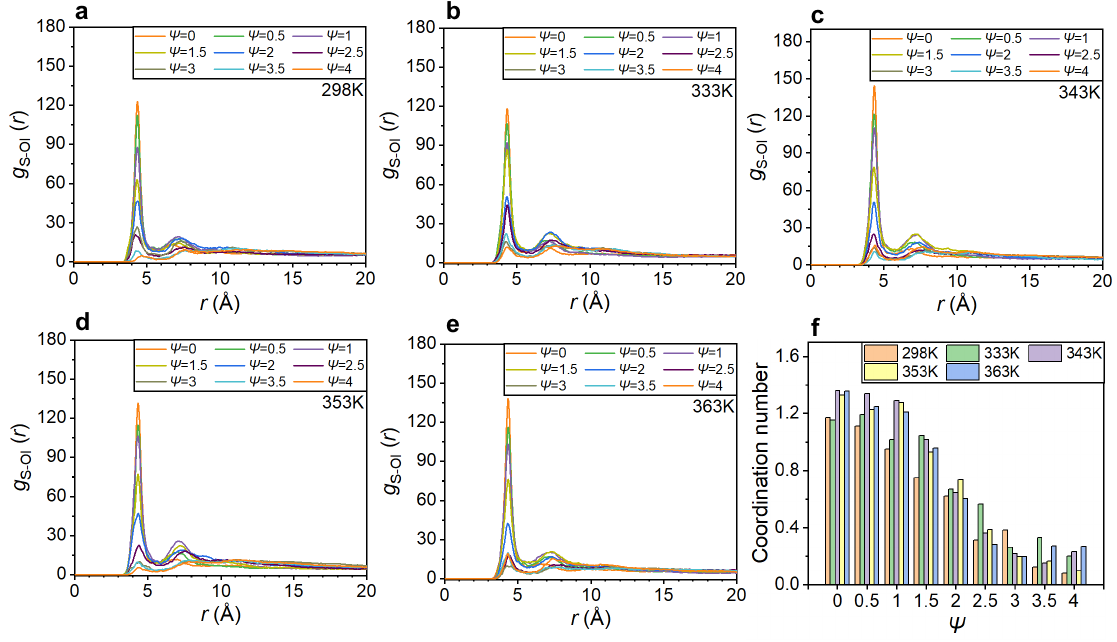}
  \caption{(a-e) RDFs of S-OI pairs at different temperatures and ammonia contents. (f) CN of the S-OI. }\label{fig:11}
\end{figure}

\begin{figure}
  \centering
  \includegraphics[scale=0.8]{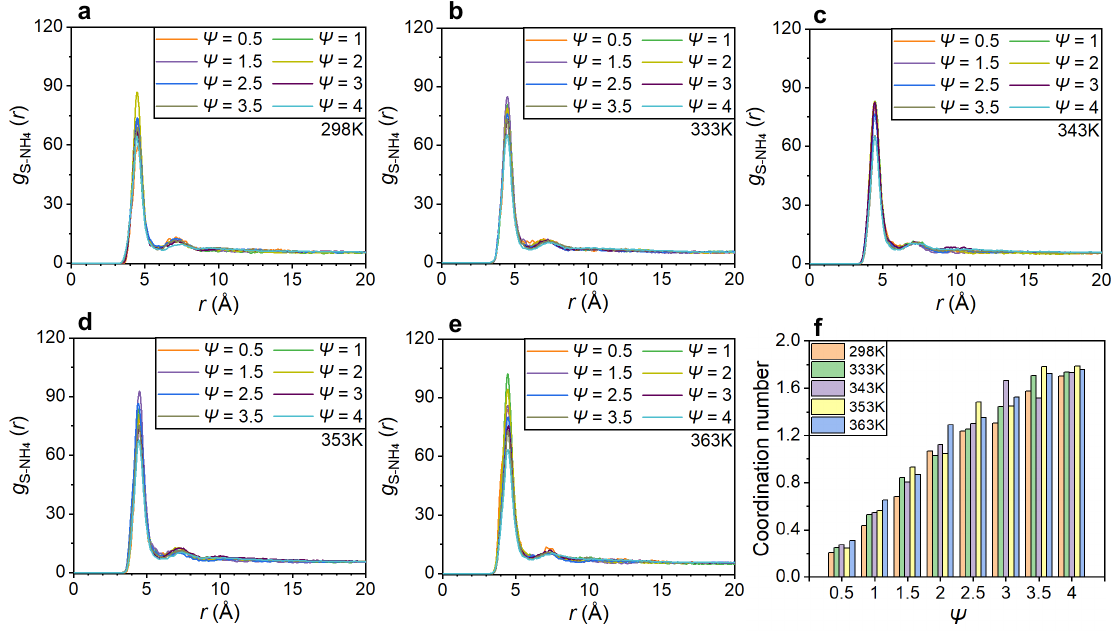}
  \caption{(a-e) RDFs of S-\ce{NH4} pairs at different temperatures and ammonia contents. (f) CN of the S-\ce{NH4}.}\label{fig:12}
\end{figure}

Regarding the phenomenon that ammonium replaces hydronium ions to occupy the sulfonic acid group of the side chain of the ionomer, we further analyzed the radial distribution function (RDF) and CN of the sulfonic acid group with hydronium ions (${{g}_{\text{S-OI}}}$) and ammonium ions (${{g}_{\text{S-NH4}}}$), where ``OI'' indicates the oxygen atom in the hydronium ions, ``S'' indicates the sulfur atom in sulfonic acid group, ``NH4'' indicates the nitrogen atom in ammonium ions. The results for S-OI and S-\ce{NH4} pairs are shown in Figures \ref{fig:11} and \ref{fig:12}. In Figure \ref{fig:11}(a-e), the position of the first peak of $g_\text{S-OI}$ appears at about 4.45 \AA, and the first valley appears at about 5.25 \AA. The position of the valley was used as the radius of the first solvation layer to obtain the CN of the S-OI pairs. We can find that the value of the first peak decrease with the increase in the ammonia content, which means that the correlation between sulfonic acid group with the hydronium ion is further reduced. The change in the system temperature has little effect on the number of hydronium ions around the sulfonic acid group, while the change in the CN is mainly due to the change in the ammonia content. With the increase in the ammonia content, the CN of S-OI is greatly reduced, because the increase in the ammonia content will lead to the capture of hydronium ions by amino and imino ions, and the hydronium ions can no longer be effectively transported through the charged sites on the sulfonate groups. The significant change in the CN is also consistent with the situation shown in the snapshots of Figure \ref{fig:02}. The distribution between the sulfonic acid group and the ammonium ion is shown in Figure \ref{fig:12}. In Figure \ref{fig:12}(a-e), the first peak of S-\ce{NH4} also appears at about 4.45 \AA, while the first valley appears at about 5.95 \AA. The CN was calculated based on this valley position. It can be found that the effect of the system temperature on the distribution of ammonium around the sulfonic acid group is still very subtle. With the increase in the ammonia content, the correlation between ammonium and sulfonic acid groups does not change much, but the CN increases significantly. This is mainly because with the increase in the ammonia content, the total number of ammonium ions also increases, resulting in an increase in the number of ammonium ions per unit space. In this condition, the ammonium ions will be attracted by the charged sites of the sulfonic acid, resulting in a steady increase in the CN.

\begin{figure}
  \centering
  \includegraphics[scale=0.68]{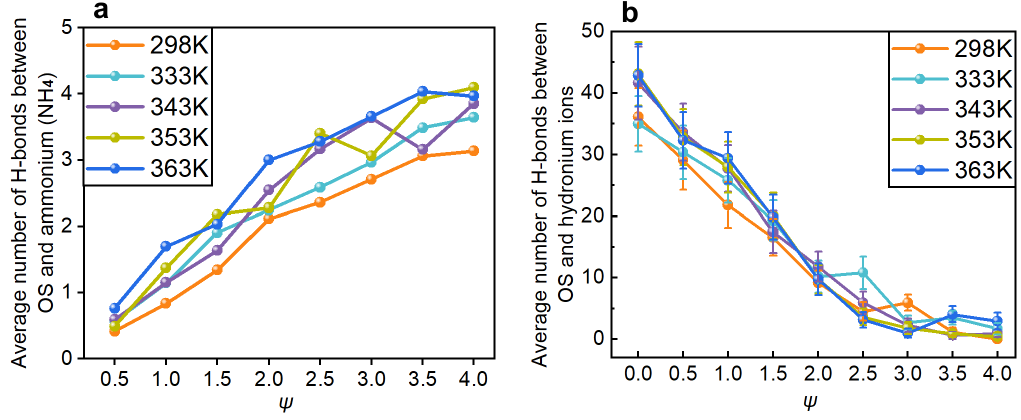}
  \caption{Average number of hydrogen bonds (a) between sulfonic acid group (\ce{SO3-}) and ammonium (\ce{NH4+}) and (b) between sulfonic acid group (\ce{SO3-}) and hydronium ion (\ce{H3O+}) at different ammonia contents ($\varphi $) and system temperatures. }\label{fig:13}
\end{figure}

To analyze the main causes of the attraction of sulfonic acid groups to hydronium ions and ammonium ions, we also counted the number of hydrogen bonds. The average number of hydrogen bonds between sulfonic acid groups and ammonium ions, and between sulfonic acid groups and hydronium ions were calculated. The oxygen atom in the hydronium ion and the nitrogen atom in the ammonium ion were set as donors, and the sulfur and oxygen atoms in the sulfonic acid group were set as the acceptors. 200 configurations in the 10-ns sampling stage were counted and averaged. As shown in Figure \ref{fig:13} (a), with the increase in the ammonia content, the number of hydrogen bonds formed between ammonium and sulfonic acid groups increases, and the increase in the system temperature will also promote the formation of hydrogen bonds between ammonium and sulfonic acid groups to a certain extent. However, it is worth noting that this is the statistic average of the total number of hydrogen bonds formed between all sulfonic acid groups and ammonium in the whole system, which is much smaller than the CN. This means that the occupation of sulfonic acid groups by ammonium ions is not through the formation of hydrogen bonds, but through charge attraction and van der Waals force. Figure \ref{fig:13}(b) shows the number of hydrogen bonds formed by hydronium ions and sulfonic acid groups. With the increase in the ammonia content, the change in the total number of hydrogen bonds is consistent with the change in the CN, showing a downward trend. When the ammonia content is less than 2.5, the hydronium ions are completely captured by the amino group and the imino group, but the hydronium ions on the outer layer of the ion clusters can still form some hydrogen bonds with the sulfonic acid group. When the ammonia content is further increased, the imino and amino ions occupy the outer layer of the ion cluster. In this condition, the hydronium ions can no longer approach the sulfonic acid group to form hydrogen bonds, and the number of hydrogen bonds is reduced to almost 0. However, in this condition, the CN of the sulfonic acid group and the hydronium ion is not 0, indicating that the attraction between them in this condition is via charge adsorption and van der Waals force.

To observe the transport behavior of ammonium ions in the ionomer layer, we selected several ammonium ions in the ionomer layer and tracked their trajectories to observe their transport behavior. Typical ammonium ion movements in the ionomer layer are provided in Figure S6 in Supplementary Material. We can see that ammonium ions are not stably adsorbed onto a single sulfonic acid group, and they rather transport through the membrane being similar to hydronium ions in the absence of ammonia ($\psi $ = 0), temporarily attached to different sulfonic acid groups over time. To quantitatively analyze ammonium ion transport, we measured the root mean square displacement of ammonium ions and calculated their diffusion coefficient, with the diffusion coefficient of hydronium ions (in the absence of ammonia, $\psi $ = 0) as a reference, as shown in Figure S3 in Supplementary Material. We can see that while ammonium ion diffusion coefficients vary slightly with the ammonia content and temperature, they remain within the same order of magnitude as those of hydronium ions. This further confirms the phenomenon observed in previous experiments \cite{uribe02, zhang09}, where introducing pure hydrogen cannot effectively or fully restore the ionomer's performance. Because ammonium ions continue to diffuse in the ionomer layer, occupying the charged sites on the sulfonic acid groups, then hindering the diffusion of hydronium ions.

\subsection{Structural and transport analysis of water clusters}\label{sec:3.4}
\begin{figure}
  \centering
  \includegraphics[scale=0.8]{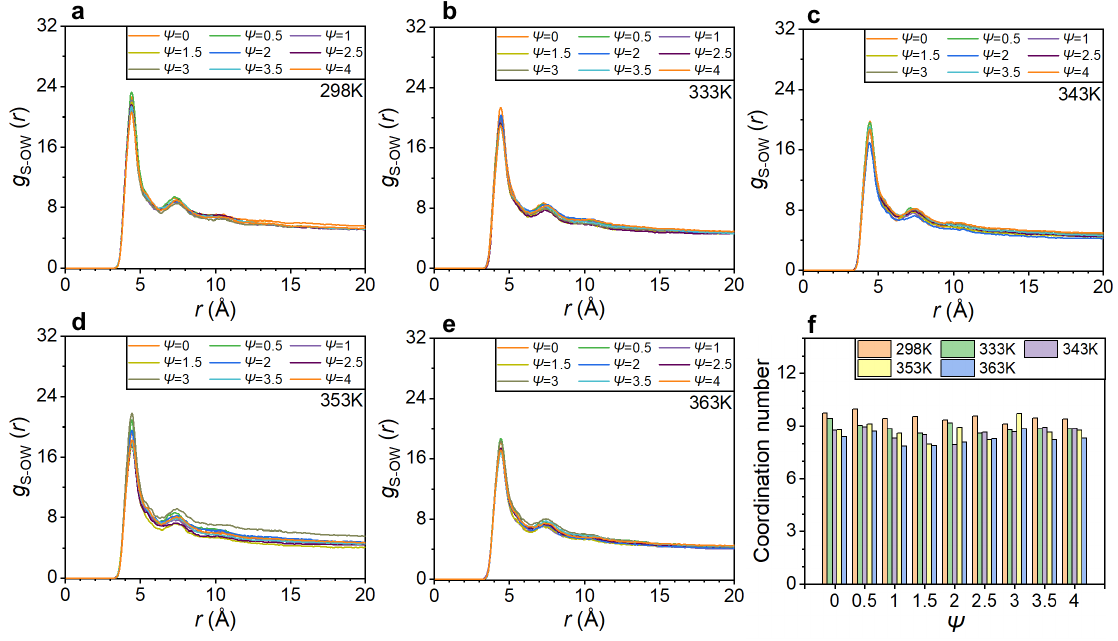}
  \caption{(a-e) RDFs of S-OW pairs at different temperatures and ammonia contents. (f) CN of the S-OW.}\label{fig:14}
\end{figure}

Since water channels provide an important route for proton transport in the ionomer layer, the morphology and distribution characteristics of the water channel are analyzed. To explore the influence of ammonia on the morphology of water channels, we calculated the RDF and CN between water molecules and sulfonic acid groups (S-OW), analyzed the surrounding situation of water molecules in the ionomer layer to sulfonic acid groups, and further analyzed the hydration state of the ionomer layer. Here, ``OW'' represents the oxygen atom in water molecules. As shown in Figure \ref{fig:14}(a-e), the position of the first valley is at about 6.25 \AA, which was used to calculate the CNs of S-OW in Figure \ref{fig:14}(f). We can find that the change in the ammonia content has little effect on the RDF and CN of S-OW pairs, indicating that the addition of ammonia will neither affect the encapsulation effect of water molecules on sulfonic acid groups, nor will it affect the correlation between water and sulfonic acid groups. With the increase in the system temperature, the CN decreases slightly. This is because the high temperature makes the water molecules more active, decreasing the stability of the interaction between water and the sulfonic acid group, and leading to the decrease in the CN.

\begin{figure}
  \centering
  \includegraphics[scale=0.8]{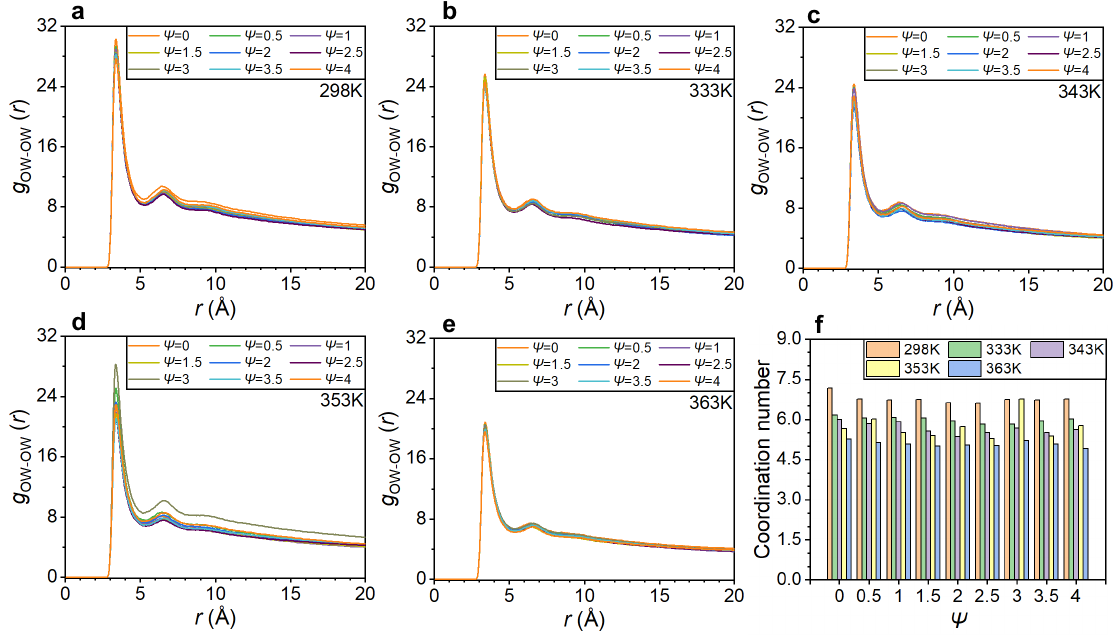}
  \caption{(a-e) RDFs of OW-OW pairs at different temperatures and ammonia contents. (f) CN of the OW-OW.}\label{fig:15}
\end{figure}

\begin{figure}
  \centering
  \includegraphics[scale=0.55]{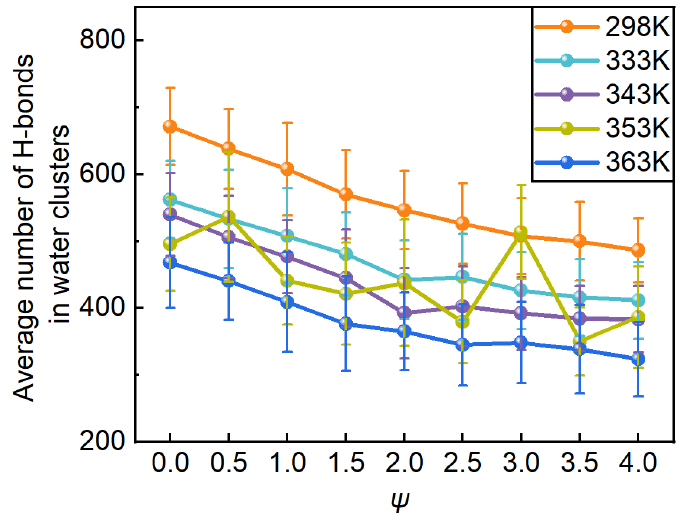}
  \caption{Average number of hydrogen bonds in water clusters at different ammonia contents ($\varphi $) and temperatures.}\label{fig:16}
\end{figure}

To further analyze the morphology of water clusters, we analyzed the RDF and CN between water molecules (OW-OW), as shown in Figure \ref{fig:15}. We also calculated the average number of hydrogen bonds in the water clusters, as shown in Figure \ref{fig:16}. In Figure \ref{fig:15}, according to the RDF curve, we set the radius of the first solvation shell layer to 5.25 \AA to calculate the CN. It can be found that the addition of ammonia has little effect on the correlation and distribution of water molecules in the water clusters, while the increase in the system temperature will reduce the correlation between water molecules. In this condition, because the activity of water itself is greatly affected by the system temperature, a higher temperature naturally leads to the decrease in the CN. Combined with the results shown in Figure \ref{fig:16}, we can find that the total number of hydrogen bonds in the water area shows downward trends with the increase in the ammonia content and also with the increase in the system temperature. Affected by the increase in the system temperature, the correlation between water molecules decreases, and the number of hydrogen bonds decreases naturally. As the ammonia content increases, ammonia and its derivatives diffuse in water channels and promote the formation of ion clusters. The ion clusters will be stuck inside the pore of the ionomer layer, thereby reducing the connectivity between water clusters. This will hinder the formation of hydrogen bonds between different water clusters, so the total number of hydrogen bonds also shows a downward trend. Furthermore, ammonia ion clusters dissociate and compress the ionomer structure, improving the homogeneity of the ionomer layer and enhancing the connectivity of water channels in the areas that is free of ammonia ion clusters, which results in the fluctuations in the number of hydrogen bonds. To further illustrate the influence of ammonia on water channels, we calculated the connectivity of water channels and the diffusion coefficient of water molecules, as shown in Figures S4 and S5 in Supplementary Material. The calculation methods for water channel connectivity and diffusion coefficient of water molecules are detailed in the Supplementary Materials. Figure S5 presents the connectivity of water channels under various temperatures and ammonia contents. With the increase in the ammonia contents, the connectivity of water channels decreases continuously, indicating that ammonia and its derivatives inhibit the correlation between water clusters. With increasing the temperature, the connectivity of water channels also decreases, likely due to the temperature-sensitive nature of hydrogen bonds. At higher temperatures, the correlation between water molecules weakens, reducing hydrogen bonds and thereby reducing connectivity. Figure S4 shows the variation in the diffusion coefficient of water molecules under various temperatures and ammonia contents. A slight decrease in the diffusion coefficient is observed with the increase in the ammonia content, while temperature elevation leads to a more substantial increase in the diffusion coefficient, correlating with the reduction in hydrogen bonds and connectivity in water channels observed earlier.

\begin{figure}
  \centering
  \includegraphics[scale=0.8]{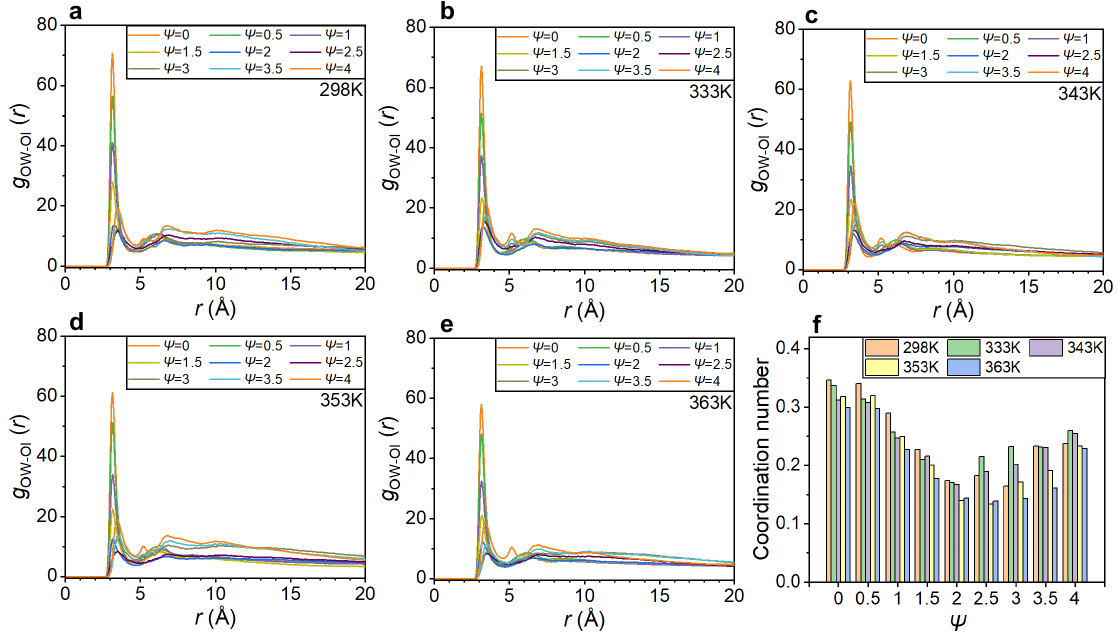}
  \caption{(a-e) RDFs of OW-OI pairs at different temperatures and ammonia contents. (f) CN of the OW-OI.}\label{fig:17}
\end{figure}

\begin{figure}
  \centering
  \includegraphics[scale=0.8]{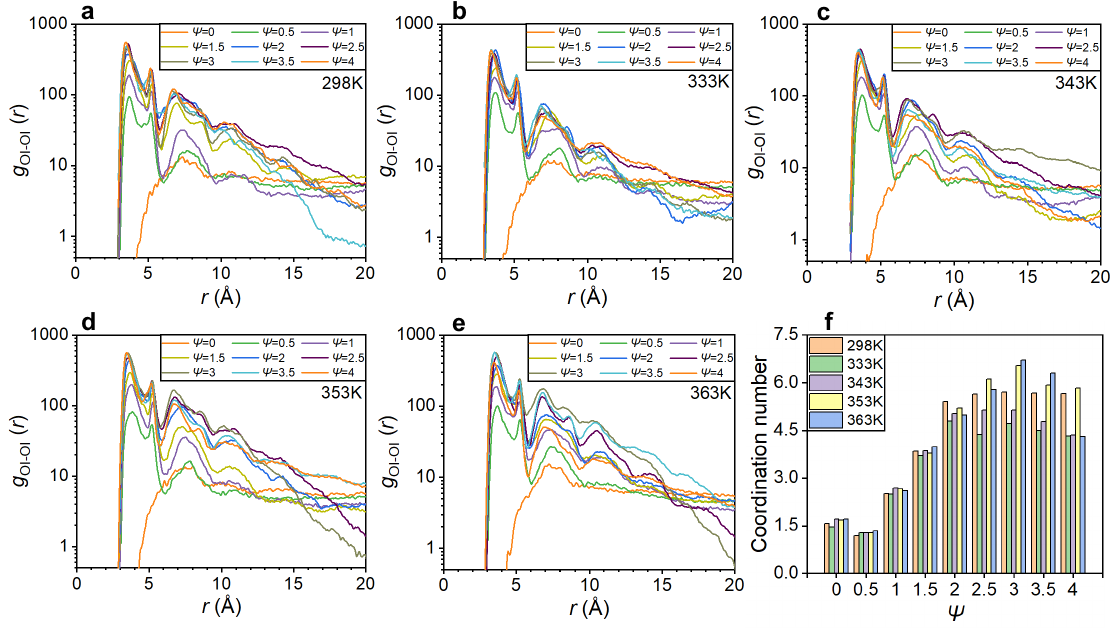}
  \caption{(a-e) RDFs of OI-OI pairs at different temperatures and ammonia contents. (f) CN of the OI-OI.}\label{fig:18}
\end{figure}

To further analyze the correlations between water molecules and hydronium ions, and between hydronium ions and hydronium ions in the water clusters, we calculated the RDFs and CNs of OW-OI and OI-OI as shown in Figure \ref{fig:17}--\ref{fig:18}. In Figure \ref{fig:17}(a-e), we can find that the change in the ammonia content and the system temperature will affect the correlation between water and hydronium ions. With the increase in the ammonia content and the system temperature, the peak value of the first peak decreases, which means that the correlation between water and hydronium ions decreases. Meanwhile, because hydronium ions are captured to form ion clusters after adding ammonia, the position of the first valley is different under different ammonia contents. When there is no ammonia in the system, the radius of the first shell layer is 4.35 \AA. When ammonia is added, the radius increases to 4.65 \AA, which is then used to calculate the CN. We can find that at the same ammonia content, the increase in the system temperature leads to the decrease in the CN, which is in line with the expectation due to the influence of molecular activity. With increasing the ammonia content, the CN decreases first and then increases. The CN reaches the lowest value when the ammonia content is about 2--2.5. Combined with the above analysis of ion clusters, we can find that under this ammonia content, the stability of ion clusters reaches the maximum, and all hydronium ions are captured by amino and imino ions. The hydronium ions are in the inner side of the cluster and the outer layer of the cluster is surrounded by a large number of amino and imino ions. In this condition, for the hydronium ions are difficult to reach the water molecules, resulting in a significant decrease in the CN. When the ammonia content is further increased, the stability of ion clusters decreases, or even decomposes into multiple ion clusters, resulting in an increase in the possibility of the contact between hydronium ions and water molecules, which leads to a further increase in the CN.

Regarding the correlation between hydronium ions, the radius of the first solvation layer also changes with the existence of ammonia, while the temperature change has little effect on the correlation between hydronium ions. When there is no ammonia in the system, hydronium ions are evenly distributed in the ionomer layer. The position of the first peak is at about 7.95 \AA, and the position of the first valley is at about 8.95 \AA, which means that about one or two hydronium ions will appear in the range of 9 \AA with the absence of ammonia. After adding ammonia, hydronium ions are captured by ion clusters, and the distance between hydronium ions is greatly reduced. The first peak appears at about 3.65 \AA, and the first valley appears at 4.85 \AA. The CN is calculated according to different positions of the first valley. With a small amount of ammonia, the hydronium ions will gradually form small ion clusters, but the CN will decrease because of the dispersion of the ion clusters. When the ammonia content increases, the free ion clusters combine to form larger ion clusters. The correlation between the hydronium ions in the clusters continues to increase, and the CN also increases accordingly. The increased correlation between hydronium ions will further enhance the stability of ion clusters, which further proves that when the ammonia content reaches about 2, the stability of ion clusters reaches the maximum. When the ammonia content further increases, the hydronium ions are all captured, and the ion clusters will gradually dissociate, the correlation between the hydronium ions in different ion clusters reduces again, and the CN also shows a downward trend. The correlation between the hydronium ions in the ion clusters is also affected by the ammonia content. When the ammonia content is 2, the correlation reaches the maximum, which further promotes the stability of the ion clusters.

\section{Conclusions}\label{sec:4}
In this study, all-atom MD simulations are performed to study the poisoning mechanism of ammonia at different ammonia contents and temperatures in CCLs of PEM fuel cells. The snapshots of the CCLs were analyzed to explore the behavior of ammonia and its derivatives in the ionomer layer. The results show that ammonium replaces hydronium ions to occupy the charged sites of the sulfonic acid groups on the side chains of the ionomers, while amino and imino ions capture hydronium ions to form ion clusters. The RDF and CN of NH-OI, \ce{NH2}-OI and NH-\ce{NH2}, along with the hydrogen bond analysis and cluster analysis, are carried out to study the transport and distribution of the ion clusters. The correlation between ions in the ion clusters changes with the increase in the ammonia content. When the ammonia content reaches 2.5, the correlation reaches the highest value, the stability of the ion cluster structure is the strongest, and the ability of ion clusters to capture hydronium ions is also the strongest. The radius of gyration and the size of ion clusters gradually increases with the increase in the ammonia content. The stability of ion clusters is also affected by the system temperature. As the temperature increases, large ion clusters will dissociate into medium-sized clusters. The main reason for the formation of the ion cluster is the hydrogen bonds between ions. The pore size distribution and snapshots analysis, along with the RDF, CN and hydrogen bond analysis, are used to study the ionomer structure and the interactions between ionomer and ammonia. Ion clusters affect the structure of ionomers. Large ion clusters can block the ionomers and hinder proton transport. Ion clusters can also squeeze the ionomers and reduce the homogeneity of the ionomer layer. The medium-sized ion clusters after dissociation affected by the system temperature can squeeze the ionomer from multiple directions, which can improve the homogeneity of the ionomer again and improve the proton transport efficiency. The adsorption of sulfonic acid group by ammonium is due to the attraction of van der Waals force and electrostatic interaction, and does not generate many hydrogen bonds. Then, the RDF, CN and hydrogen bond analysis were performed to explore the structure and transport of water clusters. High temperature significantly reduces the stability of hydrogen bonds between water molecules, reduces the number of hydrogen bonds between water clusters, and reduces the connectivity of water clusters. The understanding of the poisoning mechanisms of ammonia provides theoretical guidance for enhancing the anti-poisoning performance of the catalyst layer in PEM fuel cells.

\section*{Acknowledgements}
This work was financially supported by the Department of Science and Technology of Inner Mongolia Autonomous Region (Grant No.\ 2022JBGS0027) and the National Natural Science Foundation of China (Grant Nos.\ 51920105010 and 51921004).

\end{document}